\documentclass[twocolumn,showpacs,showkeys,footinbib,superscriptaddress]{revtex4}
\usepackage{graphicx}
\usepackage{amsmath,amsfonts,amssymb,bm,color}
\begin{document}

\title{Measurement of helium-3 and deuterium stopping power ratio for
negative muons}

\author{V.M.~Bystritsky}
\affiliation{ Laboratory of Nuclear Problems, Joint Institute for Nuclear
Research, Dubna 141980, Russia}
\author{V.V.~Gerasimov}
\affiliation{ Laboratory of Nuclear Problems, Joint Institute for Nuclear
Research, Dubna 141980, Russia}
\author{J.~Wo\'zniak}
\affiliation{Faculty of Physics and Applied Computer Science,
AGH University of Science and Technology, PL--30059 Krakow, Poland}

%\date{\today}

\begin{abstract}

The measurement method and  results measuring of the stopping
power ratio of helium-3 and deuterium atoms for muons slowed down
in the D/$^3$He mixture are presented. Measurements were performed
at four values of pure $^3$He gas target densities, $\varphi_{He}
= 0.0337,\ 0.0355,\ 0.0359,\ 0.0363$ (normalized to the liquid
hydrogen density) and at a density 0.0585 of the D/$^3$He mixture.
The experiment was carried out at PSI muon beam $\mu$E4 with the
momentum P$\mu =34.0$ MeV/c. The measured value of the mean
stopping ratio $S_{^3\!He/D}$ is $1.66\pm 0.04$. This value can 
also be interpreted as the value of mean reduced ratio of probabilities  
for muon capture by helium-3 and deuterium atoms.

\end{abstract}


\maketitle
%{\bf \large
%%Measurement of stopping power ratio for $\mu^-$ of helium and deuterium atoms
%in D/$^3$He mixture}
%Measurement of helium and deuterium stopping power for $\mu^-$
%}
%%%%%%%%%%%%%%%%%%%%%%%%%%%%%%%%%%%%%%%%%%%%%%%
%
\section{Introduction}
Atomic capture of pions and muons stopped in a mixture of hydrogen and helium
isotopes has been a subject matter of quite a lot of experimental
\cite{petru76,bystr83,bubak86,budya67,banni83,kottm87,bystr93,kottm93,bystr91}
and theoretical investigations
\cite{koren75,koren80a,koren80b,cohen83,koren87,
dolin89,koren91,koren92,koren96,fesen96,cohen04}.
The investigation of this process is important for understanding
physics of exotic system and also for studying elementary processes occurring
when negatively charged particles stop in a material.
To separate two processes -- —atomic capture and transfer of
muons from $\mu$-atoms of hydrogen isotopes in the course of their de-excitation
to helium nuclei -- —is practically impossible if x-rays are used, as usually,
as diagnostics.
What one usual observes experimentally is the result of interference
of a few processes accompanying the muon capture by atoms of
hydrogen and helium isotopes.
Therefore, it is quite a problem to extract unambiguous information on the
law of initial capture of muons by atoms of the hydrogen (deuterium)--helium
mixture components.
The earlier made assumption that the probability for direct muon capture by
atoms of the mechanical H$_2$(D$_2$)/M (M is $^{3,4}$He, Ar, Ne ...) mixture
is
proportional to the charge and concentration of each component (Fermi--Teller
$Z$-law \cite{fermi47}) turned out to be wrong.
Experimental data \cite{koren92}  revealed deviation from the
Fermi--Teller law.
The $Z$-law is based on the assumption that the atomic capture probability is
proportional to muon energy loss on atoms of the mixture components.
Actually, there is no simple relation between the stopping power of a
particular type of atom and the probability for muon capture by this atom.

For the binary mixture He/H it is convenient to express the atomic
capture probalities $W_{He},W_H$ by per-atom capture ratio $A$
(reduced ratio) defined  as $(W_{He}/C_{He})/(W_H/C_{H})$
\begin{equation}
W_H=\frac{1}{1+Ac},\ \
W_{He}=\frac{Ac}{1+Ac},
\label{eq1bb}
\end{equation}
where $c=C_{He}/C_H$ is the ratio of the helium and hydrogen
atomic concentrations.

The authors of Ref. \cite{petru76}, who measured the capture
probabilities of ($\pi^-$)-mesons by hydrogen and helium-3 atoms
in the H/He mixture, have found that expressions (\ref{eq1bb}) fit
well the experimental data when the slowing-down parameter $S$ is
used instead of the capture ratio $A$
%
%In \cite{petru76}
%it has been found that the expressions (\cite{eq1bb}) fit well the
%experimental data when to use the slowing down parameter $S$ instead of the
%capture ratio $A$
%such
%expressions for capture probabilities of ($\pi^-$)-mesons by hydrogen and
%helium atoms in the H/He mixture which fitted well the
%experimental data were adopted
%
\begin{equation}
W_H=\frac{1}{1+Sc},\ \
W_{He}=\frac{Sc}{1+Sc}.
\label{eq1}
\end{equation}
Here $S = \overline{s}_{He}/\overline{s}_{H}$ is the ratio of the
averaged stopping powers,  where
\begin{equation}
s_i={\mathcal A}_i\left ( -\frac{dE}{d\xi}\right )_i, \ \
i =\  \mathrm{He,\ H},
%S =\left
%( \overline{\frac{dE}{dx}}\right )_{He} \left ( \overline
%{\frac{dE}{dx}}\right )_{H}^{-1} = 1.84\ \pm 0.09.
\label{eq2}
\end{equation}
are per-atom stopping powers of $^3$He and H
atoms (expressed in $\mathrm{MeV}\cdot \mathrm{cm}^2/\mathrm{atom}$),
$\xi$ is the mass thickness and ${\mathcal A}_i$ are the atomic masses.

Formulae (\ref{eq1}) can also be used for negative muons captured
in the H/He mixture (with a proper $S$-value) due to similarity of
$\mu^-$ and $\pi^-$ masses (207$m_e$ and 273$m_e$, respectively).

Recently a series of experiments on the study of $\mu$-atomic and
$\mu$-molecular processes in a D/$^3$He mixture has been carried
out at the Paul Scherrer Institute meson factory
\cite{bystr04,bystr05,bystr06}. In the experiments we measured the
following characteristics: the nuclear fusion rate in a
charge-asymmetrical muon complex (d$\mu^3$He); the probability for
transition of the d$\mu$ atom from the excited to the ground state
($q_{1s}$); intensities of x-ray radiation of $\mu$He atoms
resulting both from muon capture by $^3$He atoms and from transfer
of the muon from the d$\mu$ atom under its de-excitation to the
$^3$He nucleus.

For correct interpretation of the above-mentioned experimental
data it was necessary to have information on the probability for
direct capture of muons by deuterium and helium atoms in the
D/$^3$He mixture. On the one hand, the value of the capture ratio
$A$ averaged over the data of the papers
\cite{petru76,bystr83,bubak86,budya67,banni83,kottm87,bystr93,kottm93,bystr91}
can be used; on the other hand, an attempt may be made to get
experimental information on $S=S_{He/D}$ as a ratio of stopping
powers of helium and deuterium atoms in an independent way from an
additional experiment. The description of the method and analysis
of the results of such measurement of the $S$-value for muons as
the ratio of stopping powers of helium and deuterium per atom is
the aim of this work.

%%%%%%%%%%%%%%%%%%%%%%%%%%%%%%%%%%%%%%%%%%%%%%%%%%%%%%%%%%%%%%%%%%%%%%
\section{Measurement method}
\label{sect1}

\begin{figure}[ht]
\includegraphics[width=0.8\linewidth]{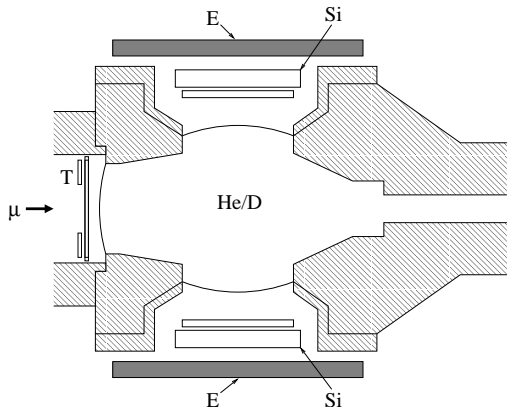}
                \caption{Scheme of the experimental setup. E -- electron
counters, Si -- silicon telescopes.
}
\label{fig:target}
\end{figure}

In our experiment we used two types of gas targets; one was a D
$+\ 5\%^3$He mixture with the density $\varphi_{mix}=0.0585$
(hereafter all atom number densities $\varphi$ are normalized to
the liquid hydrogen density, $n_o=4.25\cdot 10^{22}$ cm$^{-3}$),
the other was a target with pure helium-3. A set of different
helium densities $\varphi_{He}$ was considered.
The deuterium-helium mixture and the helium targets were exposed
to the same muon beam in order to keep the same initial energy
distributions of muons entering the targets. The momentum of the
muonic beam was chosen such as to stop all entering muons inside
the D/$^3$He target. In the final stage of the slowing-down muons
are captured by the target atoms, form muonic atoms and finally
decay via the \ \ $\mu^-\rightarrow e^-+\nu_\mu+\overline{\nu}_e$\
\ reaction. Decay electrons are then markers of the stopping
events.
A schematic view of the experimantal setup is shown in Fig.
\ref{fig:target} (see \cite{bystr04,borei98} and the next section
for details).

\begin{figure}[ht]
\includegraphics[width=\linewidth]{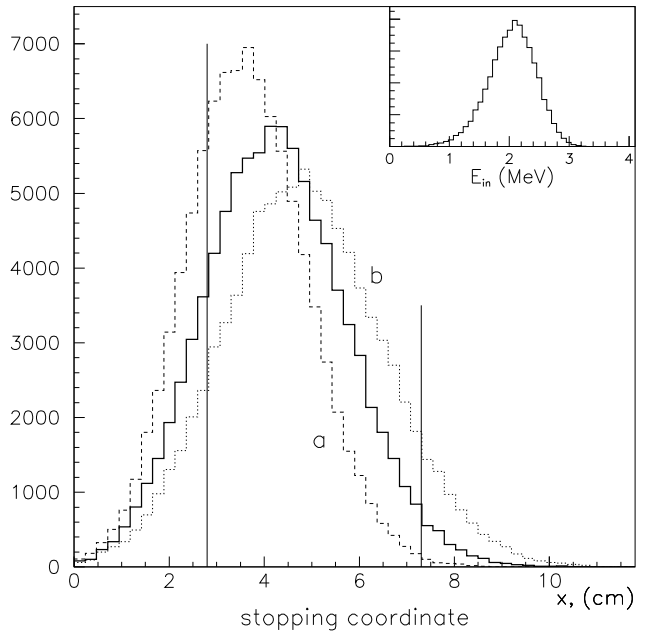}
                \caption{Stopping distributions calculated for three targets
with different densities $\varphi$: D/$^3$He mixture
($\varphi_{mix} =0.0585$, thick solid line), $^3$He($\varphi_{He}
=0.0320$, histogram (a)), $^3$He($\varphi_{He} =0.0380$, (b)). The
beam momentum is 34.0 MeV/c. The energy of incident muons,
$E_{in}$, is shown in the insert. } \label{fig:example4}
\end{figure}

Varying the gas density in the pure helium target we changed the
position of the maximum in the muon stopping distribution along
the target length.
The example of the Monte Carlo simulations presented in Fig.
\ref{fig:example4} illustrates such a situation. The stopping code
\cite{jacot97} with Ziegler parameterization of the stopping
powers \cite{ziegl85} was  used for obtaining the muon stop
distributions in the targets. Calculations were performed for the
34.0 MeV/c muon beam. Stopping distributions were calculated for
three targets: the D$ + 5\%^3$He mixture with the density
$\varphi_{mix}=0.0585$ (thick solid line), pure $^3$He with
density $\varphi_{He}=0.0380$ (dashed line) and
 pure $^3$He
with the density $\varphi_{He}=0.0330$ (dotted line). Vertical
lines show the position of the central (spherical) part of the
target; this part is directly seen by the electron counters. The
stopping coordinate $x$ is a distance to the stopping point taken
from the entrance window along the beam direction. It is clear
that the number of electrons detected by the electron counters
will change when the stopping distribution is shifted.

\begin{figure}[ht]
\includegraphics[width=0.48\linewidth,angle=90]{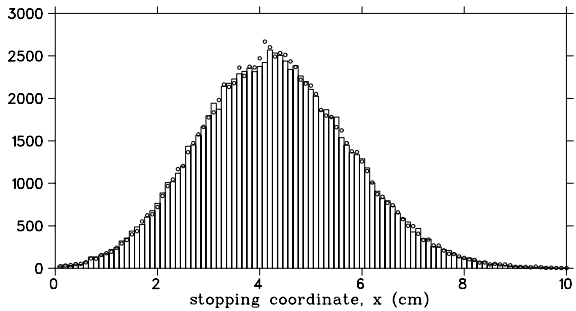}
                \caption{Calculated stopping distribution
for the D/$^3$He target (histogram, solid line) and the equivalent
distribution for pure the $^3$He target with the density
$\tilde{\varphi}_{He}=0.0342$ (open circles). }
\label{fig:distrib}
\end{figure}

The density of the pure helium target can be chosen such that the
number of muon stops detected via decay electrons is the same as
in the D$_2$ + $^3$He mixture (we denote this density by
$\tilde{\rho}_{He}$). In such a case spatial distributions of muon
stops are equivalent. Such equivalency of the stopping
distributions was verified by performing the MC simulations with a
set of different $^3$He densities and examining the differences by
$\chi^2$-analysis for both distributions (from D/$^3$He and $^3$He
targets). Figure \ref{fig:distrib} shows, as an example, the
distributions where the minimum of $\chi^2$/df = 0.92 was achieved
for pure helium density $\tilde{\varphi}_{He}=0.0342$. Similar
equivalence of the stopping distributions is obtained in the plane
perpendicular to the beam axis. The details of the experimental
results discussed in the next section also justify the assumption
of equivalence of spatial distributions of muon stops for a
selected He target density.
%
%The similar equivalence is in the plane perpendicular to yhe beam axis.
%Such behavior of the stopping distributions is ilustrated on
%%Fig. \ref{fig:distrib} where the MC simulated spectra are shown. A set of
%simulations was performed with the different $^3$He densities
%case.
%
%%% Figure: distrib %%%

From the identity of the spatial distributions in both targets (considering that the initial muon energy distributions are also identical) follows
equality of the ranges of stopped muons for any initial muon
energy $E_{in}$
\begin{equation}
\int^{E_{in}}_0\frac{1}{(-\frac{dE}{dx})_{He}}dE =
  \int^{E_{in}}_0\frac{1}{(-\frac{dE}{dx})_{mix}}dE.
\label{eq10}
\end{equation}
The above equation can be rewritten in terms of the atomic
stopping powers $s_{He},\ s_D$
\begin{equation}
\int^{E_{in}}_0\!\!\frac{dE}{\tilde{\varphi}_{He}\,s_{He}} =
  \int^{E_{in}}_0\!\!\frac{dE}{\varphi_{mix}(S^{-1}C_D+ C_{He})\,s_{He}},
\label{eq10a}
\end{equation}
where $S$ is the ratio of stopping powers being the subject of the
measurement
\begin{equation}
S = \frac{s_{He}(E)}{s_D(E)}\,.
\label{eq10b}
\end{equation}

In appendix \ref{sect:append}  we argue how a simple and more
useful formula for the mean ratio of the stopping powers can be
derived from relations (\ref{eq10}),(\ref{eq10a}) when the
behavior of the individual stopping powers of helium-3 and
deuterium is taken into account. Such a formula reads
\begin{equation}
S_{He/D}=\overline{S}=S(\overline{E})= \frac{C_D
\varphi_{mix}}{\tilde{\varphi}_{He} - C_{He} \varphi_{mix}},
\label{eq11f}
\end{equation}
where $\overline{E}$ is the average energy of the initial muon
energy distribution. The above formula gives the recipe for
measurement of the mean ratio of the helium-3 and deuterium atomic
stopping powers.

The specific density $\tilde{\varphi}_{He}$ (needed for obtaining
$S$ by formula (\ref{eq11f})) is experimentally established by
measuring the yields of electrons from muon decays in the D/$^3$He
mixture target and in a set of pure $^3$He targets.
%As the muon decay electron detection efficiency
%depends upon the coordinates of the muon stopping point (in the gas volume or
%in the target walls), it can be stated that muon stops in $^3$He and in the
%D/$^3$He mixture may be equal in number only if distributions of muon stops
%over the target volume are identical in both cases.
%%%%%%%%%%%%%%%%%%%%%%%%%%%%%%%%%%%%%%%%%%%
\section{Measurement and results}
\label{exp} The experiment was carried out at muon channel
µ$\mu$E4 at the Paul Scherrer Institute meson factory. An
experimental setup (see Fig. \ref{fig:target}) developed for
studying the muon-catalysed nuclear fusion reaction
$d\mu^3\mathrm{He} \rightarrow \alpha+\mu+p\ (14.6\ \mathrm{MeV})
$ \cite{bystr06} was used to measure $S_{He/D}$.
%The setup is described in detail in \cite{bystr04}.
%

The body of the cryogenic gas target was made of pure $Al$ in the
form of a sphere 250 cm$^3$ in volume. There were five kapton
windows 55 to 135 $\mu$ thick in the target body. The entrance
window for the muon beam was 45 mm in diameter, its kapton was
pressed with a stainless steel flange with a 1-mm-thick gold ring
inserted in it. The other four windows were arranged in a circle
and were intended for detection of charged products of fusion
reaction in the $d\mu^3\mathrm{He}$ reaction and muonic X-rays
from $\mu$He µatoms. Electrons from the decay of muons stopped in
the target were detected by four pairs of plastic scintillator
counters (E) installed around the target.
%Protons from reaction (16) and
%mu-X-ray radiation of   K?-lines of Heµ atoms were detected by three silicon ( )
%telescopes and a 0.17-cm3 GeS detector respectively.

%There is an important point worth mentioning.
%Since the muon decay electron detection
%efficiency depends upon the coordinates of the muon stop point (in the gas
%volume or in the front and rear target walls), the shift of the maximum of
%the space distribution function of
%muon stops in the target along the muon beam direction could be "traced"
%(??)
%by variation in the density of the target with pure $^3$He (with the muon
%beam momentum remaining unchanged).
The experiment included four runs with pure $^3$He and one run
with the D/$^3$He mixture. The experimental conditions are given in
Table \ref{tab:1}.
\begin{table}[ht]
\begin{ruledtabular}
      \caption{Conditions of the experiment. $C_{He}$ is the atomic
concentration of helium, $N_{\mu}$ is the number of muons that
entered the target.} \label{tab:1}
\begin{tabular}{ccccccccccc}
Run &Target &Temp. &Presssure &$\varphi$ &$C_{He}$ &$N_{\mu}$\\
&& [K] & [atm] & [LHD] & [\%] & [$10^9$]\\
\hline
1 & 3He      & 32.9 & 6.92 & 0.0363 & 100 & 1.3625\\
2 &&&                 6.85 & 0.0359 &&      0.7043\\
3 &&&                 6.78 & 0.0355 &&      0.7507\\
4 &&&                 6.43 & 0.0337 &&      0.4136\\
\hline
5 & D/$^3$He & 32.8 & 5.11 & 0.0585 & 4.96 & 8.875\\
\end{tabular}
\end{ruledtabular}
\end{table}

Information on the distribution of muon stops in the target volume
in experiments with pure $^3$He and with the D/$^3$He mixture can
be gained from analysis of time distributions of muon decay
electrons in $^3$He and D/$^3$He targets
\begin{equation}
\frac{dN_e^{He}}{dt}=B_{Al}e^{-\lambda_{Al}t}+B_{Au}e^{-\lambda_{Au}t}+
B_{He}e^{-\lambda_{He}t}+B,
\label{eq30}
\end{equation}
\begin{equation}
\frac{dN_e^{D/^3\!He}}{dt}=F_{Al}e^{-\lambda_{Al}t}+F_{Au}e^{-\lambda_{Au}t}+
F_{D/^3\!He}e^{-\lambda_{D/^3\!He}t}+F,
\label{eq31}
\end{equation}
where $B_{Al},\ B_{Au},\ B_{He},\ F_{Al},\ F_{Au}$ and
$F_{D/^3He}$ are normalized amplitudes, $B$ and $F$ are the levels
of accidental coincidences, and $\lambda_{Al},\ \lambda_{Au},\
\lambda_{He},\ \lambda_{D/^3He}$ are the rates of muon
disapperance in the target wall material and in the target gas.

Measuring the normalized partial amplitudes
$B_{He}$ and $F_{D/^3He}$ and knowing the muon decay
electron detection efficiency averaged over the muon energy distribution, we
can determine the number of muon stops in $^3$He and the D/$^3$He mixture.

Comparing the results of the analysis of the data obtained under different
experimental conditions we introduced a quantity $R$ for convenience.
$R$ is a ratio between the number of electrons from decays of muons
stopped in $^3$He (or in D/$^3$He mixture) and the number of incident muons
\begin{equation}
R=\frac{N_e}{N_\mu}.
\label{eq31a}
\end{equation}
The scenario of the experiment was as follows. First, the target
was filled with a D/$^3$He mixture at the density
$\varphi_{mix}=0.0585$ and the initial muon beam momentum $P_\mu$
was varied to find its value corresponding to the maximum $R$.
This value of $R$ corresponded to the highest density of muon
stops in the gas at the center of the target. Then the target was
filled with pure $^3$He to the pressure at which the number of
stops in the target approximately corresponded to the number of
muon stops in the D/$^3$He mixture or, to be more exact, at which
the numbers $N^{D/^3He}_e$ and $N^{He}_e$ of detected electrons
from decays of muons stopped in the gas (in $^3$He and D/$^3$He
mixture) were rather close. The $^3$He pressure in the filled
target was selected using the program for calculation of muon
energy loss on the passage of the muons through various materials
\cite{jacot97}.

To find out how the ratio $R$ varies with the $^3$He density, four runs were
carried out with the helium density varying in the interval 1\% --–7\% from
run to run (see Table \ref{tab:1}).

\begin{figure}[ht]
\includegraphics[angle=90,width=\linewidth]{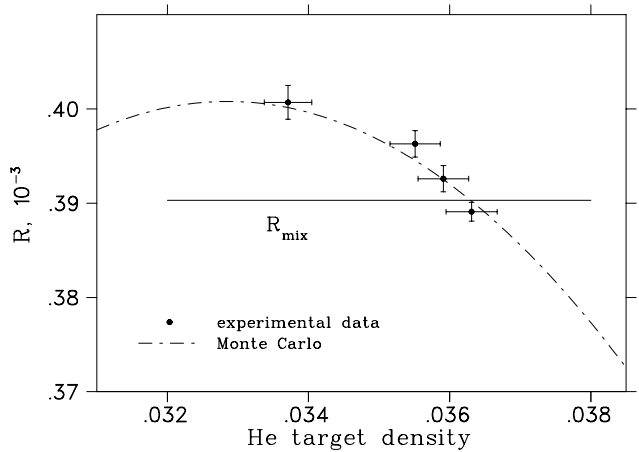}
                \caption{$R$=${N_e}/{N_\mu}$ is the ratio between the number of
decay electrons detected by the electron counters and the number
of incident muons  (black circles) for four pure $^3$He targets
used in the experiment (densities: $\varphi=0.0337,\ 0.0355,\
0.0359,\ 0.0363$). The horizontal solid line represents the
experimental value of $R$ for the D/$^3$He target (mixture of
deuterium and 5\% of $^3$He, $\varphi=0.0585$). The dash-dotted
line is the Monte Carlo simulation normalized by the factor
${R_{exp}(\varphi_{mix})}/{R_{MC}(\varphi_{mix})}$. MC
calculations were performed using theoretical helium stopping
powers scaled by the factor of 0.96 in order to obtain better
description of the experimental points. } \label{fig:R}
\end{figure}

%This helium density variation range was chosen to meet the following
%requirement: the target thickness variation should be significantly smaller
%than the muon beam momentum spread in equivalent units of muon range.
%In the experiment under discussion this requirement was met.
%In this case an approximately linear character of the $R-\varphi$ relation
%should show up.
%
%%%%%%%%%%%%%%%%%%%%%%%%%%%%%%%%%%%%%%%%%%%%%
%\section{Analysis}
%\label{anal}
%
%
\begin{table*}[ht]
\begin{ruledtabular}
      \caption{Ratio $R$ measurements for runs 1 -- 5. $N_\mu$ is the number
of muons entering the target and $N_e$ is the number of detected
electrons from muon stops in the $^3$He and D/$^3$He targets.}
\label{tab:2}
\begin{tabular}{cccccc}
Run & Target & $\varphi$ & $N_\mu$ & $N_e$ & $R$  \\
&& [LHD] & [$10^9$] & [$10^6$] & [$10^{-3}$] \\
\hline
1 & 3He & 0.0363     & 1.3625 & 0.5302 (14) & 0.3891 (10) \\
2 &     & 0.0359     &  0.7043 & 0.2765 (10) & 0.3926 (14) \\
3 &     & 0.0355     &  0.7507 & 0.2975 (10) & 0.3963 (14) \\
4 &     & 0.0337     &  0.4136 & 0.1657 (8) & 0.4007 (18) \\
\hline
5 & D/$^3$He & 0.0585 &  8.875 & 3.4635 (35) & 0.3903 (4) \\
\end{tabular}
\end{ruledtabular}
\end{table*}
To find the number of muon stops in runs with $^3$He and the D/$^3$He mixture,
time distributions of muon decay electrons were approximated by expressions
(\ref{eq30}) and (\ref{eq31}) (see ref. \cite{bystr06} for more detailed
description).
Table \ref{tab:2} presents the numbers of detected muon decay
electrons  as well as the ratios $R=N_e/N_\mu$ measured in runs
1--5 with pure helium and D/$^3$He mixture. The results of the
measurement are also shown in Fig.~\ref{fig:R}, where four
experimental points of $R$ for pure helium  versus target density
are plotted. The abscissa of the intersection point of
$R(\varphi_{He})$-dependence with the horizontal line representing
the $R(\varphi_{mix})$ value for the D/$^3$He mixture is just the
one to determine the value of the equivalent helium density
$\tilde{\varphi}_{He}$.

In order to evaluate the character of $R$-dependence on
$\varphi_{He}$ the auxiliary Monte Carlo calculations were
performed for our experimental conditions. The result of the
simulation (dash-dotted line in Fig. \ref{fig:R}) shows the
non-linearity of the R-dependence.
%
%and have only auxiliary
%significance.
%
%Presented MC calculations performed for our experimental
%conditions have only auxiliary significance to show the non-linearity of the
%R-dependence and to help
%
%%% Figure: R %%%

From the analysis of the data presented in Fig. \ref{fig:R} the
equivalent helium density was found:
$\tilde{\varphi}_{He} = 0.0363 \pm 0.0005.$
According to formula (\ref{eq11f}), the experimental value of the
stopping power ratio of helium-3 and deuterium atoms is
\begin{equation}
S_{He/D}=1.66 \pm 0.04.
\label{eq35}
\end{equation}
%

%%% Figure: R %%%
%\begin{figure}[ht]
%\includegraphics[angle=90,width=0.95\linewidth]{R.ps}
%                \caption{$R=\frac{N_e}{N_\mu}$ is the ratio of
%decay electron numbers detected by the electron counters and muon stop
%numbers in the
%target (open circles) for four pure $^3$He targets used in the experiment
%(densities:
%$\varphi=0.0337,\ 0.0355,\ 0.0359,\ 0.0363$). Horizontal solid line represents the
%experimental value of $R$ for ``lpmix'' target (mixture of deuterium and 5
%\% of $^3$He, $\varphi=0.0585$). Black circles show the Monte Carlo
%simulations normalized by factor $\frac{R_{exp}(lpmix)}{R_{MC}(lpmix)}$.
%Simulated
%value of $R$ is defined as $R^{MC}=\frac{N_e^{MC}}{N_\mu^{MC}}$, where
%$N_\mu^{MC}$ is the total number of muons entering the target and $N_e^{MC}$ is
%the number of muon stop events belonging to the central
%part of the target ($2.8 < z < 7.3$ cm). Beam momentum is 34.0 MeV/c.
%For condition $R_\mathrm{He} =R_{mix}$ fulfilled
%($\varphi_\mathrm{He}\approx 0.0353$) the value of $\overline{A}$
%(see Eq. (\ref{eq11f})) from MC results is:
%$\overline{A}=\frac{C_D
%\varphi_{mix}}{\varphi_\mathrm{D}-C_\mathrm{He}\varphi_{mix}}=1.72$.
%}
%\label{fig:R}
%%
%\end{figure}
%
%

As was mentioned earlier, the fact of the identity of the muon
stop distributions for the D/$^3$He target at $\varphi_{mix}$ and
the He target at $\tilde{\varphi}_{He}$ is crucial for our
analysis. There are noteworthy points confirming this fact:
\par ({\it i})   The numbers of muon stops in the entrance ring of
$Au$ and the target walls of $Al$ are equal (per incident muon) in
both cases;\par ({\it ii})   The ratios between the numbers of
stops in the target walls and the gas are equal in both cases:\par
$N_\mu^{Al}(\tilde{\varphi}_{He})/N_\mu^{He}(\tilde{\varphi}_{He})\!=
\!N_\mu^{Al}(\varphi_{mix})/N_\mu^{D/^3He}(\varphi_{mix}), $
$N_\mu^{Au}(\tilde{\varphi}_{He})/N_\mu^{He}(\tilde{\varphi}_{He}) =
N_\mu^{Au}(\varphi_{mix})/N_\mu^{D/^3He}(\varphi_{mix}). $
%
%%%%%%%%%%%%%%%%%%%%%%%%%%%%%%%%%%%%%%%%%%%%%%%%%%%%%%%%%%%%%%%%
%\section{Concluding remarks}
%
%
%In Appendix \ref{sect:appendb} we discuss the result of our measurement
%from the point of view of the muon capture by deuterium and helium atoms.
%

An additional remark concerning the interpretation of our
measurement can be made. It is clear that the muon slowing-down
and subsequent atomic capture are closely connected to each
another. A discussion presented in Appendix \ref{sect:appendb}
shows that such connection is so close  that allows us to enlarge
the interpretation of the measurement result from the point of
view of the atomic muon capture.
%
%In Appendix \ref{sect:appendb} we argue that the result of our measurement
%can be interpreted from the point of view of the muon capture in
%deuterium and helium atoms.
%
%Namely,
%the founded value of $\overline{S}$ represents also the reduced ratio
%of muon capture probabilities by helium and
%deuterium atoms.
%
%This value of \vmb{$\overline{S(He/D)}$} agrees within statistical
%errors with the measured values of this quantity treated as a
%ratio between specific ionization loss of muons per He and
%deuterium atom.
Concluding, we state that the result of the experiment in question
is a  ratio $S_{He/D}$ of specific stopping powers of helium and
deuterium atoms while, on the other hand, this quantity
 is phenomenologically equal to a reduced
ratio of probabilities, $A$, for muon capture by helium and deuterium
atoms.

%%%%%%%%%%%%%%%%%%%%%%%%%%%%%%%%%%%%%%%%%%%%%%%%%%%%%%%%%%%%%%%%%%
\appendix
\section{Mean ratio of stopping powers. }
\label{sect:append}

\begin{figure}[ht]
\includegraphics[angle=90,width=0.78\linewidth]{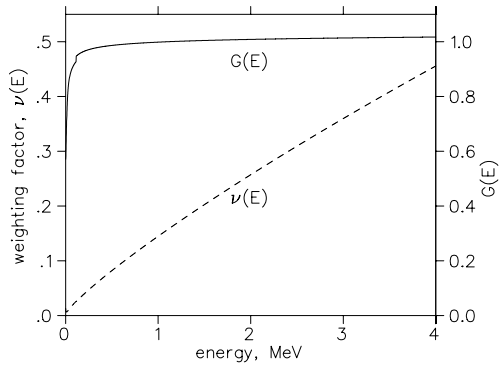}
                \caption{The functions $\nu (E)$ and $G(E)$ (formulae
(\ref{eq11b}) and (\ref{eq11c}), respectively). $G(E)$ is calculated for
the densities $\tilde{\varphi}_{He}=0.0342,\ \varphi_{mix}=0.0585$.
}
\label{fig:g}
\end{figure}

Using notations
\begin{equation}
\nu (E) = \frac{1/s_{He}(E)}{1/\int^{E_{in}}_0 1\, /s_{He}(E)\, dE},
\label{eq11b}
\end{equation}
\begin{equation}
G(E) =
\frac{\tilde{\varphi}_{He}}{\varphi_{mix}(S(E)^{-1}
C_D+C_{He})},
\label{eq11c}
\end{equation}
we can rewrite the equality of ranges (\ref{eq10a}) as
\begin{equation}
\int^{E_{in}}_0\nu(E)\, G(E)\, dE = \overline{G} = 1\, .
\label{eq11a}
\end{equation}
Two components of the integrated function in Eq. (\ref{eq11a})
depend on the muon energy $E$ in a quite different manner as is
seen in Fig. \ref{fig:g}.

%
%%% Figure: g %%%
%
%%% Figure: GG %%%
\begin{figure}[ht]
\includegraphics[angle=90,width=0.88\linewidth]{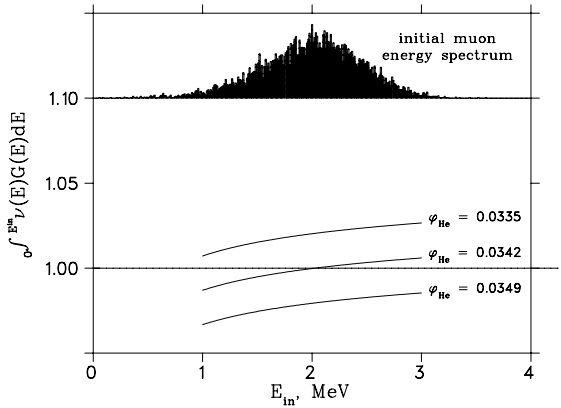}
                \caption{Ratio of muon ranges in D/$^3$He mixture and pure
helium-3 targets $\overline{G}=\int^{E_{in}}_0\nu(E)\, G(E)\, dE$
calculated for different initial muon energies $E_{in}$, for three
helium-3 target densities: 0.0335, 0.0342, 0.0349. The real energy
spectrum of muons entering the targets is also shown (top). }
\label{fig:integral}
\end{figure}
The weighting function $\nu (E)$ (normalized to unity) is strongly
energy dependent and decreases roughly linearly with decreasing
energy. $G(E)$ (energy dependent via $S(E)$) is, contrary to
$\nu(E)$, approximately constant in a wide energy region. In Fig.
\ref{fig:integral} the calculated ratio of the muon ranges in
D/$^3$He and pure helium-3 targets
$\overline{G}=\int^{E_{in}}_0\nu(E)\, G(E)\, dE$ as a function of
the initial muon energy, $E_{in}$, is presented. For given
densities $\varphi_{mix},\ \varphi_{He}$, this ratio is
practically independent of $E_{in}$. It is a consequence of the
behaviour of the integrated function (or, in other words, due to
similar dependence of deuterium and helium stopping powers on the
muon energy).
In the energy interval
1.8 -- 2.3 MeV (50 \% of beam muons belong to this interval) the
relative change of $G$  is 0.4 \%, and for interval 1.3 -- 2.7 MeV
(90 \% of muons) the respective change is 1 \%.

The quantity $\overline{G}$ in Eq. (\ref{eq11a}) represents the
ratio of the ranges of muons with the initial energy $E_{in}$ in
D/$^3$He mixture and in pure $^3$He targets. Basically, equality
(\ref{eq11a}) is fulfilled for a given energy $E_{in}$ for
especially chosen $\tilde{\varphi}_{He}$ (as is seen in Fig.
\ref{fig:integral}). For another energy the other density
$\tilde{\varphi}_{He}$ should be, in principle, adjusted. But as can
be seen from Fig. \ref{fig:integral}, such uncertainty in
$\tilde{\varphi}_{He}$ is very small (less than 1 \% in the range
of our muon energy spectrum) and can be neglected.

In view of the above considerations it is reasonable to use an
approximation
\begin{equation}
\overline{G}\ \approx G(\overline{S(E)}) \approx G(S(\overline{E})),
\label{eq11d}
\end{equation}
where $S(\overline{E})$ is the ratio of the atomic stopping powers
taken for the average energy of the initial muon spectrum $\sim$2
MeV. Then from eqs. (\ref{eq11a}) and (\ref{eq11d}) follows the
equality
\begin{equation}
\frac{\tilde{\varphi}_{He}}{\varphi_{mix}(\overline{S}\,^{-1}C_D+C_{He})}=1,
\label{eq11e}
\end{equation}
and finally formula (\ref{eq11f}).
%

%%%%%%%%%%%%%%%%%%%%%%%%%%%%%%%%%%%%%%%%%%%%%%%%%%%%%%%
%\appendix
\section{Atomic capture}
\label{sect:appendb}
Slowed-down muons are captured by the target atoms and finally end
their life decaying in the atomic orbit \footnote{A muonic atom is
formed in an excited state, then the muon fast cascades to the
ground state. The muon transfer to another type of atom is also
possible during the cascade but such transfer has no significance
for the electron detecting. Only primary atomic capture is crucial
for the appearing of decay the electron which is for us an
indicator of the muon stop. The cascade also does not change the
coordinate of the stopping point because the cascading time is
negligibly small.}, \footnote{A small part of muons can survive
the atomic capture without decay in the atomic orbit. It is a case
when the muon is freed in the result of the chain reaction with
$d\mu ^3He$ molecule formation and subsequent nuclear fusion. Such
muons can be stopped again or escape the target. In any case, such
rare events can be neglected in our considerations.}.

An alternative phenomenologocal interpretation of our measurement
is the following. Each muon stop in the target gas is followed by
an atomic capture either on a deuterium atom or on a helium atom.
We identify the stopping event with an atomic capture event
because we detect the decay electrons appearing after the muon is
captured (electron decay in flight can be neglected as the slowing
down time is about 1 ns, the probability of such decay is $\sim
10^{-4}$). The number of decay electrons, which appear as the
consequence of muon stop at a given point of the target, depends
on the density of atoms and on the probability of the atomic muon
capture.

Let $N_\mu$ muons enter the target with a given initial energy
distribution. Through a volume $\Delta V$ located at any given
point of the target there pass $\Delta N_\mu$ muons. A part of
them, $\Delta N_{s}$, having a sufficiently low energy due to
slowing-down, are captured in $\Delta V$ and produce decay
electrons. These muons which are not captured in $\Delta V$
continue to slow down and are stopped at another point of the
target. The number of decay electrons (i.e. the number of stopped
and then captured muons) in $\Delta V$ is proportional to the
number of entering muons $\Delta N_\mu$ and to the muon capture
probability described by the capture cross section $\sigma_c$.
%averaged over the energy spectrum of muons that
%stopped in the ($D_2+^3He$)-mixture and were captured by $^3He$
%and D atoms.
The mean muon capture probability in $\Delta V$ is
$\overline{\Sigma}_c \Delta x$, where $\overline{\Sigma}_c =
n(He)\overline{\sigma}_{He}+n(D)\overline{\sigma}_{D}$ is the
macroscopic capture cross section, $n$ are the densities of helium
and deuterium atoms, $\overline{\sigma}_{He}$ and
$\overline{\sigma}_{D}$ are the muon capture cross sections for
helium and deuterium atoms, respectively (averaged over the muon
energy), and $\Delta x$ is thickness of the volume $\Delta V$
along the direction of the muon beam.
Under the assumption that the muon atomic capture cross sections
$\sigma_{He}(E),\ \sigma_{D}(E)$ are the same in both cases
(monoatomic or mixture target), and do not depend on the He
concentration, the respective numbers of the captured muons  can
be written as
\begin{equation}
\Delta N_{stop}^{He} =
\Delta N_\mu  n_o\varphi_{He} \overline{\sigma}_{He} \Delta x,
\label{eq19}
\end{equation}
\begin{equation}
\Delta N_{stop}^{mix} =
\Delta N_\mu  n_o\varphi_{mix} (C_D\overline{\sigma}_{D} +
             C_{He}\overline{\sigma}_{He}) \Delta x,
\label{eq20}
\end{equation}
for a pure helium-3 target and for a target with a D+$^3$He
mixture, respectively.

For a chosen density of the pure helium target,
$\tilde{\rho}_{He}$, the spatial stopping distribution is the same
as for the mixture target (such a situation is illustrated in Fig.
\ref{fig:distrib}). Then, for the same volumes $\Delta V$ in both
targets, the number of stops in helium  the target with density
$\tilde{\rho}_{He}$ is equal to the number of stops in the mixture
with the density $\rho_{mix}$
\begin{equation}
\Delta N_{stop}^{He} = \Delta N_{stop}^{mix}.
\label{eq20a}
\end{equation}

Since the above local equality is valid for any point of the target
we can rewrite it
(using eqs. (\ref{eq19}) and (\ref{eq20})) as a general
relation
%, having the same form as is given by Eq. (\ref{eq17})
%
%\begin{equation}
%\frac{n_o \tilde{\varphi}_{He}}{N_a}p(He) = \frac{n_o \varphi_{mix}}{N_a}
%(C_D p(D) + C_{He} p(He)).
%\label{eq21}
%\end{equation}
%
\begin{equation}
\tilde{\varphi}_{He} A =   \varphi_{mix}(C_D + A C_{He}),
\label{eq22}
\end{equation}
where
\begin{equation}
A = \frac{\overline{\sigma} (He)}{\overline{\sigma} (D)}
\label{eq23}
\end{equation}
is the ratio of the probabilities of muon capture by helium-3 and
deuterium atoms.
Rewriting Eq. (\ref{eq22}) in the form
\begin{equation}
A = \frac{C_D \varphi_{mix}}{\tilde{\varphi}_{He} - C_{He}
\varphi_{mix}},
\label{eq24}
\end{equation}
one obtains the formula for the quantity $A$, which is equivalent to
Eq. (\ref{eq11f}).

From the above considerations and taking, in particular,
into account the identity of the relations (\ref{eq11f}) and
(\ref{eq24}), one can state that  the value of the atomic stopping
power ratio $\overline{S}$, which we measure in our experiment, is
also (under indicated restrictions) a value of the reduced
probability ratio, $A$, for the muon capture by helium-3 and
deuterium atoms.

\acknowledgements
%%%%%%%%%%%%%%%%%%%%%%%%%%%%%%%%%%%%%%%%%%%%%%%%%%%%%%%%%%%%%%%
We are thankful to
P.~Knowles, F.~Mulhauser, L.A.~Schaller, H.~Schneuwly,
V.F.~Boreiko, W.~Czaplinski, M.~Filipowicz,
V.N.~Pavlov, C.~Petitjean, and V.G.~Sandukovsky for very useful discussions and for
their help during the data taking period.
This work was supported by the Russian Foundation for Basic Research,
Grant No.~01--02--16483, the Polish State Committee for Scientific
Research, the Swiss National Science Foundation, and the Paul Scherrer
Institute.

%%%%%%%%%%%%%%%%%%%%%%%%%%%%%%%%%%%%%%%%%%%%%%%%%%%%%%%%%%%%%%

\end{document}